
%
%
%

\input phyzzx
\REF\CK{
Callan C G and Klebanov I R 1994 {\it Phys.Rev.Lett.}{\bf 72} 1968;
\hfill\break
Callan C G, Klebanov I R, Ludwig A W W and Maldacena J M 1994 {\it
Nucl.Phys.}\ {\bf B422}\ 917
}
\REF\PT{
Polchinski J and Thorlacius L 1994 {\it Phys.Rev.}\ {\bf D50}\ 622
}
\REF\GZ{
Ghoshal S and Zamolodchikov A B 1994 {\it Int.J.Mod.Phys.}\ {\bf A9}\
3841;\hfill\break
Ghoshal S, 'Bound State Boundary $S$ Matrix of the sine-Gordon Model',
1993 {Preprint RU-93-51}
}
\REF\FS{
Fendley P 1993 {\it Phys.Rev.Lett.}\ {\bf 71} 2485;\hfill\break
Fendley P and Saleur H, 'Deriving Boundary $S$ Matrices', 1994
{Preprint USC-94-001};
\hfill\break
Fendley P, Saleur H and Warner N P, 'Exact Solution of a Massless
Scalar Field with a Relevant Boundary Interaction',
1994 {Preprint USC-94-010};
\hfill\break
Saleur H, Skorik S and Warner N P,
'The Boundary sine-Gordon Theory: Classical and Semi-classical
Analysis',
1994 {Preprint USC-94-013}
}
\REF\FK{
Fring A and K{\" o}berle R 1994 {\it Nucl.Phys.}\ {\bf B421}\ 1592
}
\REF\Sas{
Sasaki R, 'Reflection Bootstrap Equations for Toda Field Theory',
1993 Preprint YITP/U-93-33;\hfill\break
Corrigan E, Dorey P E, Rietdijk R H  and Sasaki R 1994 {\it
Phys.Lett.}\ {\bf B333}\ 83
}
\REF\ABBBQ{
Alcaraz F C, Barber M N, Batchelor M T, Baxter R J and Quispel G R W
1987 {\it J.Phys.A: Math.Gen.}\ {\bf 20}\ 6397
}
\REF\Che{
Cherednik I V 1984 {\it Theor.Math.Phys.}\ {\bf 61}\ 977
}
\REF\Skl{
Sklyanin E K 1988 {\it J.Phys.A: Math.Gen.}\ {\bf 21}\ 2375
}
\REF\PS{
Pasquier V and Saleur H 1990 {\it Nucl.Phys.}\ {\bf B330}\ 523;
}
\REF\MNR{
Mezincescu L, Nepomechie R I and Rittenberg V 1990
{\it Phys.Lett.}\ {\bf A147}\ 70; \hfill\break
Mezincescu L and Nepomechie R I 1991
{\it Int.J.Mod.Phys.}\ {\bf A6}\ 5231,
{\it ibid.}\ {\bf A7}\ 5657
}
\REF\DVG{
de Vega H J and Gonz{\' a}lez Ruiz A 1993 {\it J.Phys.A: Math.Gen.}\ {\bf
26}\ L519; \hfill\break
'Boundary $K$-matrices for the XYZ, XXZ and XXX Spin Chains',
{Preprint LPTHE-PAR 93/29}
}
\REF\KS{
Kulish P P and Sklyanin E K 1991 {\it J.Phys.A: Math.Gen.}\ {\bf 24}\ L435
}
\REF\Bax{
Baxter R J 1982 {\it Exactly Solved Models in Statistical Mechanics},
Academic Press, London
}
\REF\FT{
Faddeev L D and Takhtajan L A 1979 {\it Russ.Math.Surv.}\ {\bf 34}\ 11
}
\REF\Lut{
Luther A 1976 {\it Phys.Rev.}\ {\bf B14}\
2153;\hfill\break
den Nijs M P M 1981 {\it Phys.Rev.}\ {\bf B23}\ 6111
}
\REF\Sklb{
Sklyanin E K 1987 {\it Funct.Analy.Appl.}\ {\bf 21}\ 164
}
\REF\Lus{
L\" uscher M  1976   {\it Nucl.Phys.}\ {\bf B117}\ 475 }
\def\al{\alpha}
\def\sn{{\rm sn}}
\def\cn{{\rm cn}}
\def\dn{{\rm dn}}
\def\tr{{\rm tr}}
\def\sne{s_{\eta}}
\def\K#1{\mathrel{\mathop{\kern0pt K}\limits^#1}}
\def\Kpt#1{\mathrel{\mathop{\kern0pt K_{+}^{t_{#1}}}\limits^#1}}
\def\Kpp#1{\mathrel{\mathop{\kern0pt K'}\limits^#1}}
\pubnum={YITP/K-1084}
\date={August 1994}
\pubtype={\quad}
\titlepage
\title{Integrable XYZ Spin Chain with Boundaries}
\author{Takeo INAMI and Hitoshi KONNO}
\address{ Yukawa Institute for Theoretical Physics, \break
Kyoto University, Kyoto 606-01, Japan.}
\abstract{
We consider a general class of  boundary terms
of the open  XYZ spin-1/2 chain compatible with integrability.
We have obtained the  general elliptic solution of
$K$-matrix obeying the boundary Yang-Baxter equation
using the $R$-matrix of the eight vertex model and
derived the associated integrable spin-chain Hamiltonian.
}
\endpage
\sequentialequations

       1+1 dimensional integrable models with boundaries find
interesting applications in particle physics as well as condensed
matter
systems. In view of this, attempts have been made recently at the
integrable extension
of conformal field theories,\refmark{\CK, \PT} both massive and
massless integrable
quantum field theories\refmark{\GZ \sim \Sas}
and solvable lattice models\refmark{\ABBBQ \sim \KS}
to those with
boundary terms.
In the case of lattice models,
relying on the earlier work by
Cherednik,\refmark{\Che}
Sklyanin has given\refmark{\Skl}
a general framework which enables us to treat this problem on an
algebraic footing.
Especially, the general solution of integrable
boundary terms has been found
in the XXZ and the XXX Heisenberg spin chain system\refmark{\DVG}
based on his framework.
In this letter, we will work out the general solution of boundary
interactions in the case of XYZ Heisenberg spin chain system.
\par
                The Hamiltonian of
the XYZ spin-1/2 chain is given by the transfer matrix of the
eight  vertex model.\refmark{\Bax}
The eight vertex model is defined in terms of the Boltzmann
weights given by
the elliptic solution
$R(u)$ of the Yang-Baxter (Y-B) equation
$$\eqalignno{
R_{12}(u-u')R_{13}(u)R_{23}(u') = &
R_{23}(u')R_{13}(u)R_{12}(u-u').&\eqname{\ybeq}
\cr}$$
Here we regard $R(u)$ as linear operators acting  on
the tensor product of vector spaces $V\otimes V$
with $V={\bf C}v_+\oplus {\bf C}v_-$ and $R_{12}\equiv R\otimes 1,
R_{23}\equiv 1\otimes R$, etc. as those acting on $V_1\otimes
V_2\otimes V_3$, where $V_i\cong V,\  i=1,2,3$.
Setting $R(u)v_{\varepsilon_1'}\otimes v_{\varepsilon_2'}=
\sum_{\varepsilon_1, \varepsilon_2}v_{\varepsilon_1}\otimes v_{\varepsilon_2}
R(u)_{\varepsilon_1 \varepsilon_2}^{\varepsilon_1'\varepsilon_2'}$
and arranging the elements of $R$ in the order
$(\varepsilon_1, \varepsilon_2)=(++),
(+-), (-+), (--)$, one can express the eight vertex $R$-matrix as follows.
$$\eqalignno{
R(u)&=\left(\matrix{\sn(u+\eta)&0&0&k\ \sn\  \eta\ \sn\  u\ \sn(u+\eta)\cr
               0&\sn\ u&\sn\ \eta&0\cr
               0&\sn\ \eta&\sn\ u&0\cr
           k\ \sn\ \eta\ \sn\ u\ \sn(u+\eta)&0&0&\sn(u+\eta)\cr}\right),
&\eqname{\rmatrix}
\cr}
$$
where $\sn\ u \equiv \sn(u; k)$ is the Jacobi elliptic function of
modulus $0\leq k \leq 1$. \par
        Let ${\cal P}_{ij}$ be the transposition
operator on $V_i\otimes V_j$, i.e. ${\cal P}(x\otimes y)=y\otimes x$.
The $R$-matrix \rmatrix\ is known to have the following desirable properties.
$$\eqalignno{
&{\rm{\bf Regularity}}\quad\qquad\qquad R(0)=r(\eta){\cal P}, \qquad \qquad
r(\eta)=\sn\ \eta,
&\eqname{\reg}
\cr
&{\rm {\bf P-symmetry}}\quad\qquad
{\cal P}_{12}R_{12}(u){\cal P}_{12}=R_{12}(u), &\eqname{\psym}
\cr
&{\rm {\bf T-symmetry}}\quad\qquad
R^{t_1t_2}_{12}(u)=R_{12}(u), &\eqname{\tsymm}
\cr
&{\rm {\bf Unitarity}}\quad\qquad\qquad\
R_{12}(u)R_{12}(-u)=\rho(u)1, \quad \rho(u)=\sn^2\ \eta-\sn^2\ u
,&\eqname{\unitarity}
\cr
&{\rm {\bf Crossing\ unitarity}}\quad
R^{t_1}_{12}(u)R^{t_1}_{12}(-u-\eta)=\tilde{\rho}(u)1, \quad\cr
&\qquad\qquad\qquad\qquad\qquad\qquad
\tilde{\rho}(u)=\sn^2\ \eta-\sn^2(u+\eta).&\eqname{\crossunit}
\cr}$$
\par
      In the case of periodic boundary condition, it is known that
the Y-B equation \ybeq\ implies a commuting family of transfer matrix.
Hence the model is integrable.
\par
     We now consider the eight vertex model with boundary interactions.
Aiming at describing integrable systems with boundaries,
Sklyanin\refmark{\Skl} has introduced a pair of matrices
$K_+(u)$ and $K_-(u)$. The effects of presence of boundaries
at the left and
right ends are solely described by $K_+(u)$ and $K_-(u)$, respectively.
$K_{\pm}(u)$ are defined as the solutions to the relations
$$\eqalignno{
 R_{12}(u-u')&\K{1}_-(u)
R_{12}(u+u')\K{2}_-(u')  \cr
            &=\K{2}_{-}(u') R_{12}(u+u')
\K{1}_{-}(u)R_{12}(u-u'),&\eqname{\bybeqa}\cr
R_{12}(-u+u')&\Kpt{1}(u)
R_{12}(-u-u'-2\eta)\Kpt{2}(u')\cr
            &=\Kpt{2}(u')
R_{12}(-u-u'-2\eta)\Kpt{1}(u)R_{12}(-u+u'),&\eqname{\bybeqb}
\cr}$$
where $\K{1}_{\pm}\equiv K_{\pm}\otimes {\rm id}_{V_2}$ and
$\K{2}_{\pm}\equiv {\rm id}_{V_1}\otimes K_{\pm}.$
The equations \bybeqa\ and \bybeqb\ are called boundary Y-B equations
and $K_{\pm}(u)$ boundary $K$-matrices.
\par
        The boundary Y-B equations
imply a commuting family of transfer matrix.\refmark{\Skl}
The transfer matrix
$t(u)$, in this case, is defined using
the $K_{\pm}$ and the monodromy matrix $T(u)$ as
$$\eqalignno{
t(u)=& \tr\Bigl[ K_+(u)T(u)K_-(u)T^{-1}(-u)\Bigr],&\eqname{\transfer}
\cr}$$
where $T(u)$ is given by
$$\eqalignno{
T(u)=& R_{N0}(u)R_{N-10}(u)\cdots R_{10}(u).&\eqname{\monodromy}
\cr}$$
The trace in \transfer\ should be taken over $V_0$.
Then,
the commuting property of $t(u)$,
$$\eqalignno{
[t(u),\  t(u')]=&0 ,&\eqname{\commuting}
\cr}$$
follows from the properties of $R$ and the
boundary Y-B equations \bybeqa\ and \bybeqb.\par
   The problem is now to
solve the equations \bybeqa\ and \bybeqb\ and find
general solutions for $K_{-}$ and $K_{+}$,
using the eight vertex $R$-matrix given in \rmatrix.
It suffices to consider the first equation, because of the following fact.
Suppose $K_{-}(u)$ is a solution of the first equation, then
the function
$$\eqalignno{
K_+(u)=&K^{t}_-(-u-\eta) &\eqname{\kplus}
\cr}$$
gives the solution for the second equation.
\par
    We now proceed to solving Eq.\bybeqa.
Write $K_-(u)$ as
$$\eqalignno{
K_-(u)=&\left(\matrix{x(u)&y(u)\cr
                     z(u)&v(u)\cr}\right), &\eqname{\kminus}
\cr}$$
we have found that, out of the sixteen equations in the boundary
Y-B equation \bybeqa, only three are independent:
$$\eqalignno{
s_-vv'+s_+xv'=& s_+vx'+s_-xx', &\eqname{\byba}\cr
yz'+k\ s_-s_+zz'=& zy'+k\ s_-s_+yy', &\eqname{\bybb}\cr
S_-s_+yx'+k\ s^2_{\eta}S_-s_-zx'& +k\ s_{\eta}S_-S_+(s_-vz'+s_+xz')\cr
           =& S_+s_-yx'+k\ s^2_{\eta}S_+s_+zx'+s_{\eta}(s_-vy'+s_+xy'),
                               &\eqname{\bybc}
\cr}$$
where we set $x\equiv x(u),\ x'\equiv x(u')$, etc., and
$$\eqalignno{
s_{\eta}\equiv& \sn\ \eta,\quad
s_{\pm}\equiv \sn(u\pm u'), \quad
S_{\pm}\equiv \sn(u\pm u'+\eta). &\eqname{\sdomo}\cr
}$$
In the following, we also use the notations
$\al(u)\equiv v(u)/x(u),\  \beta(u)$
$\equiv z(u)/y(u)$ and
$\gamma(u)\equiv y(u)/x(u)$.\par
    Dividing \byba\ by $xx'$, one obtains
$$\eqalignno{
\al(u')=&{\sn(u+u')\al(u)+\sn(u-u')\over
              \sn(u-u')\al(u)+\sn(u+u') }. &\eqname{\zetaprime}
\cr}$$
Taking the limit $u'\rightarrow u$ of ${\al(u')-\al(u)\over
u'-u}$, one obtains the following differential equation.
$$\eqalignno{
{{\rm d}\al(u)\over {\rm d} u}=&-{1-\al(u)^2\over \sn\ 2u}.
&\eqname{\diffzeta}
\cr}$$
After the change of variable $t=\sn\ u$, the integration of
\diffzeta\ takes the form
$$\eqalignno{
\int{d\al \over 1-\al^2}=&-{1\over 2}\int dt{1-k^2t^4\over
 t(1-t^2)(1-k^2t^2)}.&\eqname{\intdiff}
\cr}$$
One can easily get the general solution of Eq.\intdiff,
$$\eqalignno{
{v(u)\over x(u)}=&{C\cn\ u\ \dn\ u-\sn\ u\over C\cn\ u\ \dn\ u+\sn\ u},
&\eqname{\solvx}
\cr}$$
where $C$ is an arbitrary constant.
\par
       In a similar way, from \bybb\ one gets
$$\eqalignno{
\beta(u')=&{\beta(u)+k\ \sn(u+u')\sn(u-u')\over
k\ \sn(u+u')\sn(u-u')\beta(u)+1},
&\eqname{\etaprime}
\cr}$$
and the differential equation
$$\eqalignno{
{{\rm d}\beta(u)\over {\rm d} u}=&-k\ \sn\ 2u\ (1-\beta(u)^2).
&\eqname{\diffeta}
\cr}$$
This implies the general solution
$$\eqalignno{
{z(u)\over y(u)}=&{\lambda(1-k\ \sn^2\ u)-1-k\ \sn^2\ u\over
\lambda(1-k\ \sn^2\ u)+1+k\ \sn^2\ u},
&\eqname{\solzy}
\cr}$$
with $\lambda$ being  another arbitrary constant.
\par
        Dividing \bybc\ by $yy'$, the third equation \bybc\ can be written
$$\eqalignno{
{\gamma(u)\over \gamma(u')}=&{\sne\ (s_-\al(u)+s_+)(1-k\ s_+s_-\beta(u'))
\over S_-s_+-S_+s_-+k\ \beta(u)\sne^2\ (S_-s_--S_+s_+)}.&\eqname{\wratio}
\cr}$$
Substituting \etaprime\ into \wratio\ and
replacing $\al(u)$ and $\beta(u)$ by the RHS of \solvx\ and \solzy,
one can factorize \wratio\ in the form of the ratio of the
same functions, one with the argument $u$ and the other with $u'$,
respectively.
We thus find
$$\eqalignno{
{y(u)\over x(u)}=&\mu\ {\lambda(1-k\ \sn^2\ u)+1+k\ \sn^2\ u\over
C\cn\ u\ \dn\ u+\sn\ u},&\eqname{\solyx}
\cr}$$
where $\mu$ is the third arbitrary constant.
\par
      From \solzy\ and \solyx, one can now get the ratio $z(u)/x(u)$.
It is then not difficult
to check that the above solutions satisfy all the
remaining equations, if one notes the identity
$$\eqalignno{
\sne^2\ {S_+s_+-S_-s_-\over S_-s_+-S_+s_-}
=&{S_+S_--s_+s_-\over 1-k^2S_+S_-s_+s_-}
=\sne\ \sn(2u+\eta). &\eqname{\ident}
\cr}$$
\par
    In summary, we have obtained the general solution of \bybeqa\
as $K_-(u)=$ $K(u;\xi,\lambda,\mu)$ with
$$\eqalignno{
&K_-(u;\xi,\lambda,\mu)\cr
&={1\over \sn\ \xi}
\left(\matrix{\sn(\xi+u)&\mu\ \sn 2u{\lambda(1-k\ \sn^2u)+1+k\
\sn^2u\over  1-k^2\sn^2\xi\ \sn^2u     }\cr
                \mu\ \sn2u{\lambda(1-k\ \sn^2u)-1-k\ \sn^2u\over
                            1-k^2\sn^2\xi\ \sn^2u}&\sn(\xi-u)\cr}
\right),&\eqname{\solk}\cr
}$$
where we set $C=\sn\xi/\cn\xi\ \dn\xi$ and replaced
$\mu\ \cn\xi\ \dn\xi$
by $\mu$.\foot{The solution obtained in Ref.[\DVG]
for the XYZ model are the special cases of
\solk\ associated with the special solutions $\al(u)^2=1,\ \beta(u)^2=1$ of
Eqs.\diffzeta\ and \diffeta. }
We normalize the
matrix $K_-(u)$ as $K_-(0)=1 $ for later convenience.\refmark{\Skl}
\par
    In the trigonometric limit $k\rightarrow 0$, where
$\sn\ u\rightarrow {\rm sin}\ u$, we recover the result in the case of
the six
vertex model given by de Vega and Gonz{\' a}lez Ruiz.\refmark{\DVG}
The rational limit is  obtained from the
trigonometric $K$-matrix by scaling $u\to \eta u, \xi \to\eta \xi$ and
taking the limit $\eta\to 0$.
\par
    Let us next consider the corresponding XYZ spin chain Hamiltonian.
Because of the equation \commuting, one can regard the transfer matrix
$t(u)$
as the generating function of integrals of motion of the system.
Its first logarithmic derivative implies the Hamiltonian.
$$\eqalignno{
H=&2r(\eta)\ t^{-1}(0)(t'(0)-\tr K'_+(0))\cr
 =&2r(\eta)\Bigl(\sum_{n=1}^{N-1}H_{n,n+1}+{1\over 2}\Kpp{1}_-(0)+
{\tr_0\K{0}_+(0)H_{N0}\over \tr K_+(0)}\Bigr),&\eqname{\logderiv}
\cr}$$
where the two-site Hamiltonian is given by
$$\eqalignno{
H_{n,n+1}=&{1\over r(\eta)}{\cal P}_{nn+1}R'_{nn+1}(0).&\eqname{\hnn}
\cr}$$
By a direct calculation with the $K$-matrices
$K_-(u)=K_-(u;\xi_-,\lambda_-,\mu_-)$ and  $K_+(u)=$\hfill\break
$K^{t}_-(-u-\eta;-\xi_+,-\lambda_+,-\mu_+)$ together with
the $R$-matrix \rmatrix,
one gets the following result.
$$\eqalignno{
H=&\sum_{n=1}^{N-1}((1+\Gamma)\sigma^x_n\sigma^x_{n+1}+
(1-\Gamma)\sigma^y_n\sigma^y_{n+1}+\Delta\sigma^z_n\sigma^z_{n+1})\cr
  &\qquad+\sn\ \eta(A_-\sigma^z_1 +B_- \sigma^+_{1}+C_-\sigma^-_1
+A_+\sigma^z_N
+B_+\sigma^+_{N}+C_+\sigma^-_{N}),
&\eqname{\hamilton}
\cr}$$
where
$$\eqalignno{
\Gamma=&k\ \sn^2\ \eta, \quad \Delta=\cn\ \eta\ \dn\ \eta\cr
A_{\pm}=&{\cn\ \xi_{\pm}\dn\ \xi_{\pm}\over 2\sn\ \xi_{\pm}},\quad
B_{\pm}={\mu_{\pm}(\lambda_{\pm}+1)\over \sn\ \xi_{\pm}},\quad
C_{\pm}={\mu_{\pm}(\lambda_{\pm}-1)\over \sn\ \xi_{\pm}}.&\eqname{\abc}
\cr}$$\par
     In conclusion, we have obtained the general elliptic solution of
the boundary Y-B equation for the $K$-matrices
and derived the Hamiltonian of the associated  XYZ spin-1/2 chain
with boundary terms.\par
     An immediate question is to find the ground state energy and the
excitation spectrum of the XYZ Hamiltonian we have derived.
The diagonalization of this
Hamiltonian can be  achieved by means of the generalized algebraic
Bethe ansatz by Sklyanin\refmark{\Skl} with some modification as in
the periodic boundary condition case.\refmark{\Bax, \FT} This subject is now
under investigation.\par
     It was shown using the results of Bethe ansatz type analysis
that, by tuning the X-Y anisotropy coupling ($\Gamma$ in Eq.\hamilton ),
the XYZ Hamiltonian with periodic
boundary
condition give rise to the quantum sine-Gordon theory in the
continuum limit.\refmark{\Lut}
In the case of open XYZ spin chain, it is of interest to ask whether
one can tune the coupling of boundary terms together with the $\Gamma$
so that one can derive its field theory limit.
The resulting theory is expected to be the quantum boundary sine-Gordon
theory.\refmark{\GZ, \FS}
In the limit $N\rightarrow\infty$, we have three boundary terms
proportional to $\sigma_1^x$, $\sigma_1^y$ and $\sigma_1^z$ associated
with three free parameters in $K_-$, whereas
the boundary term proposed by
Sklyanin\refmark{\Sklb} and
Ghoshal and Zamolodchikov\refmark{\GZ} has two parameters.
It is necessary to explain this difference in the analysis of the
continuum limit.
\par
     Furthermore, the higher logarithmic derivatives of commuting
transfer matrix give infinite number of conserved quantities.
In the closed XYZ spin chain case, it was shown\refmark{\Lus}
that the conservation
laws associated with these quantities yield selection rules in the
scattering process of the quantum sine-Gordon theory.
It is an interesting question to ask how the parameters appearing in the
$K$-matrices affect the scattering process of the boundary sine-Gordon
theory.
\par
We will present our result on these problems in
future publications.\par
\ack
     The authors  would like to thank Akira Fujii, Michiaki Takama,
Tetsuji Miwa for discussions. They also owe much to Ryu Sasaki and
Evgenii Sklyanin who kindly read the manuscript and gave valuable
comments. They are also indebted to Ryu Sasaki for directing their
attention to the subject of integrable models with boundaries.
This work was supported in part by
Soryushi Shogakukai, and the grant-in-aid for scientific research
on priority area 231, the Ministry of Education, Science and Culture.
\refout

\end

\bye